\documentclass[a4paper,11pt]{article}
\pdfoutput=1 

\usepackage{jheppub} 

\usepackage[T1]{fontenc} 
\usepackage{multirow}
\usepackage{multicol}

\title{\boldmath A Meta-analysis of the 8 TeV ATLAS and CMS SUSY Searches}

\author[a]{Benjamin Nachman}
\author[b]{and Tom Rudelius}

\affiliation[a]{SLAC National Accelerator Laboratory, Stanford University\\Menlo Park, CA 94025, U.S.A.}
\affiliation[b]{Jefferson Physical Laboratory, Harvard University\\Cambridge, MA 02138, U.S.A.}

\emailAdd{bnachman@cern.ch}
\emailAdd{rudelius@physics.harvard.edu}

\abstract{
Between the ATLAS and CMS collaborations at the LHC, hundreds of individual event selections have been measured in the data to look for evidence of supersymmetry at a center of mass energy of 8 TeV.  While there is currently no significant evidence for any particular model of supersymmetry, the large number of searches should have produced some large statistical fluctuations.  By analyzing the distribution of p-values from the various searches, we determine that the number of excesses is consistent with the Standard Model only hypothesis.  However, we do find a shortage of signal regions with far fewer observed events than expected in both the ATLAS and CMS datasets (at $1.65\sigma$ and $2.77\sigma$, respectively).   While not as compelling as a surplus of excesses, the lack of deficits could be a hint of new physics already in the 8 TeV datasets.
}

\begin{document} 
\maketitle
\flushbottom

\section{Introduction}

The 2012 discovery of the Higgs boson by the ATLAS~\cite{Aad:2012tfa} and CMS~\cite{Chatrchyan:2012ufa} experiments at the Large Hadron Collider (LHC) brought widespread attention to the field of high energy physics.  The discovery also reaffirmed the existence of the hierarchy problem due to the small value of the Higgs boson mass.  One popular solution to the hierarchy problem is Supersymmetry (SUSY), in which every Standard Model (SM) particle has a supersymmetric counterpart.   Since no SUSY particles have been observed, SUSY must be broken: the SUSY partner masses are much larger than their SM analogues.  However, the LHC provides a large center of mass energy which could be high enough to produce some of these SUSY particles.  Both ATLAS and CMS have conducted extensive searches for SUSY in a multitude of final states, with various numbers of jets, leptons, and photons.  The kinematic reach of the detectors have been exploited in order to be sensitive to high mass particles, which may be produced with a low cross section.  

Every search for SUSY consists of three pieces: 1) an event selection to maximize sensitivity to a particular model of interest, 2) an estimation of the number of SM events which will pass the selection for a given amount of data, and 3) a comparison of the measured number of events in the region with the predicted number.  As the data are stochastic, the last step requires a statistical analysis which incorporates the systematic uncertainties in the predicted event yields.   For a particular statistical model of the SM prediction, if the probability $p$ that the measurement could have been generated from the prediction is small, then one has evidence for SUSY.  One usually converts the value $p$ into a Gaussian equivalent  number of standard deviations and then the generally agreed upon threshold for `evidence' is $3\sigma$ and for `discovery' is $5\sigma$.  However, both ATLAS and CMS have performed many searches.  In particular, each analysis usually involves many selections and so there are hundreds of searches between the two collaborations.   Statistical fluctuations alone should then lead to several high $n\sigma$ measurements.  By studying the distribution of p-values, we analyze the compatibility of the 8 TeV ATLAS and CMS SUSY searches with the SM-only hypothesis.   The procedure is similar to the analysis of the 7 TeV SUSY searches~\cite{NachmanRudelius}, with a few additions that are discussed in the analysis and results sections.

\section{Constructing the Dataset}

Even though the 8 TeV dataset was collected in 2012, both ATLAS and CMS are continuing to analyze the data.  Most likely, searches will continue to become public until the 14 TeV run efforts are fully operational in the beginning of 2015.  Therefore, we arbitrarily cutoff the searches considered for this analysis at the SUSY 2014 conference (July 20, 2014).  This includes 17 ATLAS papers~\cite{Aad:2014nra,Aad:2014mra,Aad:2014lra,Aad:2014kra,Aad:2014yka,Aad:2014bva,Aad:2014wea,Aad:2014iza,Aad:2014pda,Aad:2014vma,Aad:2014mha,Aad:2014qaa,Aad:2014nua,Aad:2013gva,Aad:2013yna,Aad:2013ija,Aad:2013wta} and 12 CMS papers~\cite{Khachatryan:2014qwa,Khachatryan:2014doa,Chatrchyan:2014aea,Chatrchyan:2014lfa,Chatrchyan:2013mya,Chatrchyan:2013fea,Chatrchyan:2013iqa,Chatrchyan:2013xna,Chatrchyan:2013xsw,Chatrchyan:2013wxa,Chatrchyan:2013lya,Chatrchyan:2012paa}.  The difficulty in assembling the dataset for the present analysis is to understand the correlations between measurements.  The general strategy is to categorize the various searches by their selections on jets, leptons, and photons.  Two analyses which have non-overlapping requirements in the number and properties of these objects are treated as uncorrelated.  For the data, this is an excellent assumption and only breaks down in the rare case that the data in one signal region is used for the background estimate of another signal region.  If two signal regions are such that one is a subset of the other, then a decorrelation procedure is attempted in order to produce two orthogonal regions.  If the yields are $x\pm \sigma_x$ and $y\pm \sigma_y$ with $x<y$, then the decorrelated regions have yields $x\pm \sigma_x$ and $(y-x)\pm \sqrt{\sigma_y^2-\sigma_x^2}$.  In all other cases, it is not possible with the information given to determine the correlations and the signal regions in question are simply not used.  In general, if there are two analyses with an unknown correlation, the one with more signal regions is preferred unless the one with fewer regions already has orthogonal selections.  Tables~\ref{tab:cms} and~\ref{tab:atlas} give some summary information about the dataset construction given the general guidelines from above.

\begin{table}[h!]
\begin{center}
\begin{tabular}{ c c c }
	arXiv reference & Category & Note \\
	\hline
  1303.2985 & Multijets &  \begin{tabular}{@{}c@{}} Regions orthogonal; drop those with \\ $H_T>800$ GeV due to overlap with 1402.4770\end{tabular}    \\
  1402.4770 & Multijets &   \begin{tabular}{@{}c@{}} Regions orthogonal; drop those with \\ $H_T\in[500,800]$ GeV due to overlap with 1303.2985\end{tabular}    \\
  1305.2390 & Multijets &  Unknown correlation with 1303.2985 and 1402.4770: remove\\
  1311.4937 & One Lepton &  \begin{tabular}{@{}c@{}} Regions orthogonal; use the LS method \\ for uncertainties when given a choice\end{tabular} \\
  1308.1586 & One Lepton &  \begin{tabular}{@{}c@{}} Unknown correlation with 1311.4937. \\ Prefer 1311.4937 as its regions are orthogonal\end{tabular}   \\
  1212.6194 & Same sign leptons & SR6 $\subseteq $ SR3 $\subseteq $ SR4 $\subseteq $ SR1 $\subseteq $ SR0.  Drop other regions.\\
  1311.6736 & Same sign leptons & \begin{tabular}{@{}c@{}} Drop regions with $\geq 2$ b-jets due to overlap with 1212.6194. \\    Arbitrarily pick the low $p_T$ region \end{tabular}  \\
  1306.6643 & Multileptons &  \begin{tabular}{@{}c@{}} Unknown overlap with 1404.5801.  \\ Use 1404.5801 as it has more regions. \end{tabular} 
   \\
  1404.5801 & Multileptons & Regions orthogonal.\\
  1405.3886 & Multileptons & Use the two lepton OS regions only.\\
  1405.7570 & Multileptons &  \begin{tabular}{@{}c@{}} Use the two lepton OS regions only.  \\ Use signal sensitive regions (as described in the text)\end{tabular} 
   \\
  1312.3310 & Diphoton & Regions orthogonal.\\
  \hline
\end{tabular}
\end{center}
\caption{An overview of the signal regions used in the meta-analysis from 8 TeV CMS searches.}
\label{tab:cms}
\end{table}

\begin{table}[h!]
\begin{center}
\begin{tabular}{ c c c }
	arXiv reference & Category & Note \\
	\hline
1308.1841  & Multijets &\begin{tabular}{@{}c@{}} 8j80$x$b $\subseteq$ 8j50$x$b, $x\in\{0,1,2\}$. Unknown \\ 
 correlations between $M_J^\Sigma$ regions and others, drop
\end{tabular}  \\
1308.2631  & Multijets & SRA $m_{CT}(350)$ $\subseteq$  SRA $m_{CT}(300)$ $\subseteq$ $\cdots$ $\subseteq$   SRA $m_{CT}(150)$  \\
1407.0608 & Multijets & M3 $\subseteq$ M2 $\subseteq$ M1; C2 $\subseteq$ C1\\
1405.7875 & Multijets &\begin{tabular}{@{}c@{}}  2jt $\subseteq$ 2jm $\subseteq$ 2jl.  2jW $\cap$ 3j unknown, drop 2jW. \\ 
 6jt $\subseteq$ 6jm $\subseteq$ 5j and 6jl (5j $\cap$ 6jl $=\{\}$ once 6jm is removed).\\
 Drop all other regions due to unknown correlations.
\end{tabular} 
\\
1406.1122 & Multijets & \begin{tabular}{@{}c@{}} SRA2 $\subseteq$ SRA1; SRA4 $\subseteq$ SRA3.  Drop SRB.  \\  SRC3 $\subseteq$ SRC2 $\subseteq$ SRC1
\end{tabular} \\
1407.0600 & Multijets &\begin{tabular}{@{}c@{}} SR-0l-7j-C $\subseteq$ B $\subseteq $ A; Drop 4j regions due to 4j $\cap$ 7j = ?
\end{tabular} \\
1407.0583 & One Lepton & \begin{tabular}{@{}c@{}} Unknown correlations between shape fit regions, \\ consider only tN\_diag (signal sensitive regions).\\
tN\_boost $\subseteq$ tN\_med $\subseteq$ tightest tN\_diag region.\\
bCb\_high $\subseteq$ bCb\_med1.  Unknown relation \\
between bCa\_low and bCa\_med, drop low.  Unknown\\
correlation between bCd, tNbC\_mix and other regions, drop
\end{tabular} \\
1407.0603 &At Least One $\tau$ & \begin{tabular}{@{}c@{}} 1$\tau$Tight $\subseteq$ 1$\tau$Loose, 2$\tau$ GMSB $\subseteq$ 2$\tau$ nGM $\subseteq$ 2$\tau$ Incl.\\
Unknown overlap between 2$\tau$ bRPV and 2$\tau$ GM,\\
drop bRPV.  $\tau$+l bGM $\subseteq$ $\tau$+l mSUGRA.  Unknown overlap\\
 between $\tau$+l GMSB and bRPV, drop bRPV
\end{tabular}  
 \\
1407.0350 & At least two $\tau$s & \begin{tabular}{@{}c@{}} C1C1 $\cap$ C1N2 = ?, drop C1C1 \\ DS-lowMass $\cap$ DS-highMass = ?, drop lowMass
  \end{tabular}  \\
1403.4853 & Two OS Leptons&   \begin{tabular}{@{}c@{}} unknown correlation of L90,120 with 1403.5294, drop \\ unknown correlation between L110-100, drop L100 \\ H160 orthogonal, drop MVA region\end{tabular}      \\
1403.5294 & Two OS Leptons &  \begin{tabular}{@{}c@{}} Jet veto regions orthogonal to other searches, drop Zjets  \\ $m_{T2}^{150}(x)\subseteq m_{T2}^{120}(x) \subseteq WWc(x) \subseteq m_{T2}^{90}(x)$, $x\in\{SF,DF\}$ \\
Overlap of $WWb(x)$ with $m_{T2}^{90}(x)$ unknown, drop
 \end{tabular} \\
1404.2500 & Same Sign Leptons &   \begin{tabular}{@{}c@{}}  Regions orthogonal.  Drop SR3Llow/high \\ due to unknown overlap with 1402.7029\end{tabular}     \\
1403.5222 & Multileptons & SR$x$b $\subseteq $ SR$x$a, $x\in\{2,3\}$ \\
1402.7029 & Three Leptons &  Regions orthogonal except SR2$\tau$a $\cap$ SR2$\tau$b = ?, drop b
\\
1405.5086 & $\geq 4$ Leptons &  SR$x$noZb $\subseteq $ SR$x$noZa, $x\in\{0,1,2\}$  \\
 1310.3675 & Disappearing Tracks & Region inclusion by increasing $p_T$ cut\\
1310.6584 & Out-of-time & For the muon veto, inclusion by jet $p_T$\\
   \hline
\end{tabular}
\end{center}
\caption{An overview of the signal regions used in the meta-analysis from 8 TeV ATLAS searches.  Note that OS = opposite sign.  Inclusions for 1407.0583 are an approximation.}
\label{tab:atlas}
\end{table}

\clearpage
\newpage

\section{Analysis}


Once the ATLAS and CMS datasets were constructed, the expected and observed distributions of p-values were computed for both a Gaussian and a lognormal distribution of the expected number of counts (the number of counts itself is assumed to be Poisson).  A p-value was assigned to each data point according to
\begin{equation}
\mbox{p-value} = \int_0^\infty{\phi(\lambda|\mu,\sigma) P_{\geq n}(\lambda)d\lambda}.
\label{pvalueeq}
\end{equation}
Here, $P_{\geq n}$ is the probability of observing $n$ or more counts given a Poisson distribution with parameter $\lambda$,
\begin{equation}
P_{\geq n}(\lambda) = \sum_{k=n}^\infty \frac{e^{-\lambda} \lambda^k}{k!} =1- \sum_{k=0}^{n-1} \frac{e^{-\lambda} \lambda^k}{k!}.
\end{equation}
We performed a similar analysis of deficits rather than excesses in the SUSY search regions by replacing $P_{\geq n}$  in \eqref{pvalueeq} with $P_{\leq n}$: the probability of observing $n$ or less counts given a Poisson distribution with parameter $\lambda$.
The function $\phi(\lambda|\mu,\sigma)$ is the probability distribution function of the specified random variable with mean $\mu$ and standard deviation $\sigma$.  These parameters are the expected value for the number of counts ($\mu$) and the uncertainty on that value ($\sigma$).  For the Gaussian distribution,
\begin{equation}
\phi(\lambda|\mu,\sigma) = \frac{1}{N\sigma\sqrt{2\pi}} e^{-(\lambda-\mu)^2/2\sigma^2},
\end{equation}
where $N$ is a normalization constant correcting for the fact that $\lambda$ cannot be negative, and so the negative part of the distribution must be cut off.  For the lognormal distribution, whose support is $\mathbb{R}_+$, no such normalization constant is required,
\begin{equation}
\phi(\lambda|\mu,\sigma) = \frac{1}{\lambda\tilde{\sigma}\sqrt{2\pi}} e^{-(\ln{\lambda}-\tilde{\mu})^2/2\tilde{\sigma}^2},
\end{equation}
with $\tilde{\mu} := \ln{\mu^2/\sqrt{\mu^2+\sigma^2}}$, $\tilde{\sigma} := \sqrt{\ln{1 + \sigma^2/\mu^2}}$ defined so that the lognormal distribution is precisely the distribution of $Y = e^X$ for a Gaussian random variable $X$ with mean $\tilde{\mu}$ and variance $\tilde{\sigma}^2$.

One might expect the distribution of p-values defined in this way to be uniformly distributed on the interval $[0,1]$ under the null hypothesis, in accordance with the usual interpretation of p-values as the probability of observing a more significant result in precisely $p\times100\%$ of studies.  However, this intuitive understanding is only correct when the distribution is continuous \cite{Hartung}, not in the case of Poisson distribution considered here.  As a result, we first computed the expected distribution of p-values under the null hypothesis and then compared this with the observed distribution of p-values.  The expected distribution of p-values is determined by summing up the probability that each particular trial would fall into one of ten bins, $(\frac{i}{10},\frac{i+1}{10}], i = 0,...,9$,
\begin{equation}
\mbox{Pr}\left(\frac{i}{10} < \mbox{p-value} \leq \frac{i+1}{10} \right) = \int_0^\infty{d\lambda f_i(\lambda) \phi(\lambda|\mu,\sigma)},
\end{equation}
where
\begin{equation}
f_i(\lambda) = \sum_{m=0}^\infty\left[ \mbox{Pr}(X = m) \times \left\{ \begin{array}{lc}
1 & \mbox{ if } \mbox{Pr}(X \geq m) \in (\frac{i}{10},\frac{i+1}{10}]  \\
0 & \mbox{ otherwise } \end{array}   \right\} \right].
\label{expecteddisteq}
\end{equation}
Here, $X \sim$ Poisson$(\lambda)$ is the random variable measuring the number of counts, and the $\geq$ in \eqref{expecteddisteq} is replaced by a $\leq$ when computing deficits below rather than excesses above the expected signal.

Some of the studied signal regions had $0$ expected events.  There is no lognormal distribution with a mean of 0, so these regions had to be discarded in performing the lognormal analysis.  Fortunately, this only applied to seven of the CMS signal regions and none of the ATLAS ones.  However, a fairly sizable fraction had an expected mean that was very close to zero.  For these trials, it is reasonable to suspect that neither a Gaussian with a cutoff imposed at $0$ nor a lognormal will provide a good approximation to the true error distribution.  As a double check, we repeated our analysis after removing all data points with $\mu-2 \sigma < 0$ ($\approx 10\%$ for ATLAS, $30\%$ for CMS) .  The results of this second analysis did not differ qualitatively from the first, indicating that the results of the original analysis are not significantly affected by the statistical modeling of these data points.

\section{Results and Discussion}

\begin{figure}[h!]
\begin{center}
\includegraphics[width=75mm]{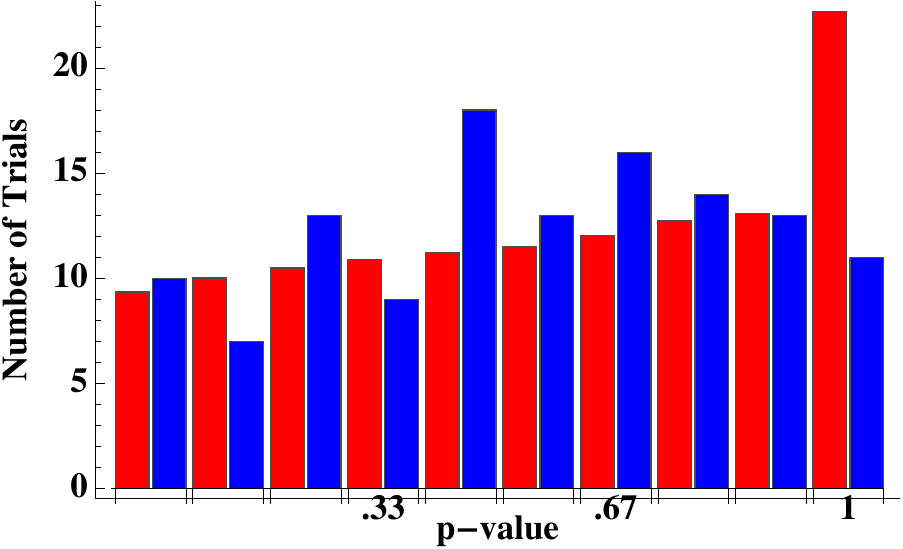}
\includegraphics[width=75mm]{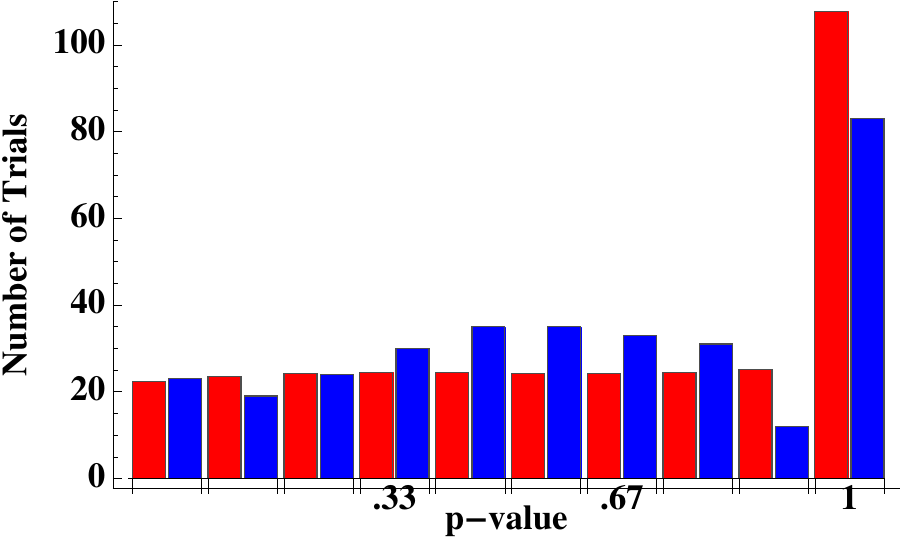}
\end{center}
\caption{Observed (blue) vs. expected (red) distribution of p-value excesses with a Gaussian error distribution for ATLAS (left) and CMS (right).}
\label{fig1}
\end{figure}

\begin{figure}[h!]
\begin{center}
\includegraphics[width=75mm]{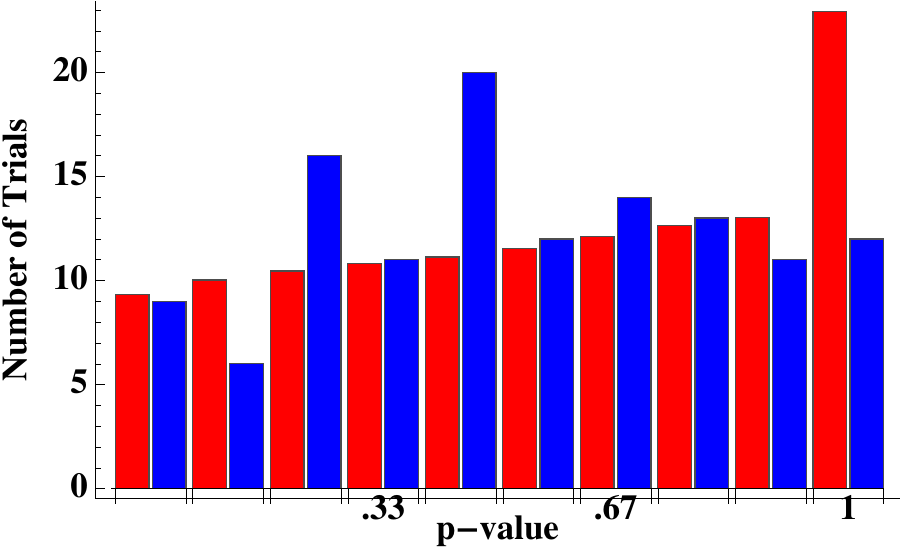}
\includegraphics[width=75mm]{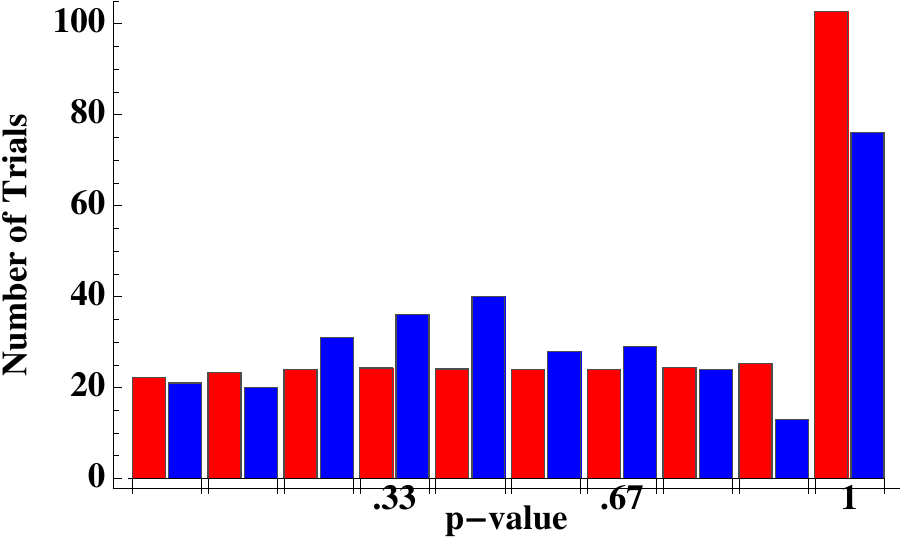}
\end{center}
\caption{Observed (blue) vs. expected (red) distribution of p-value excesses with a lognormal error distribution for ATLAS (left) and CMS (right).}
\label{fig2}
\end{figure}

\begin{figure}[h!]
\begin{center}
\includegraphics[width=75mm]{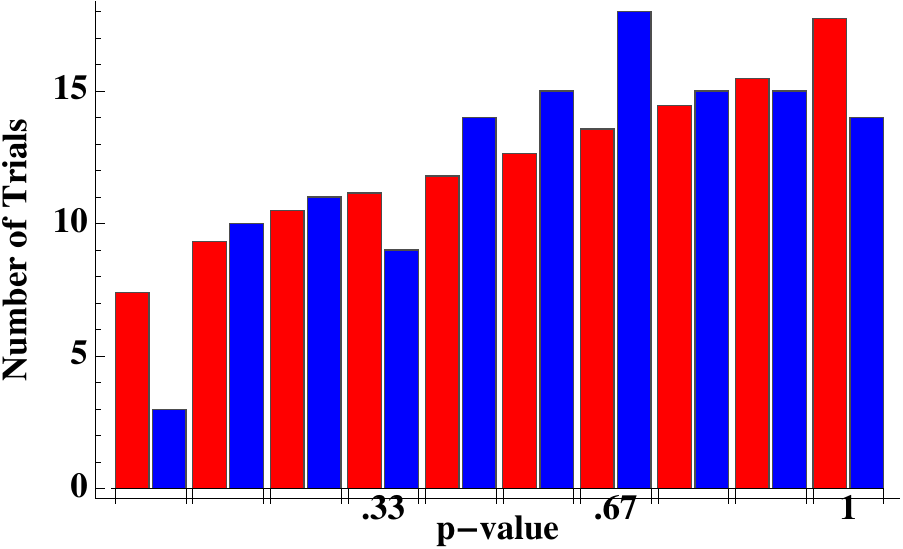}
\includegraphics[width=75mm]{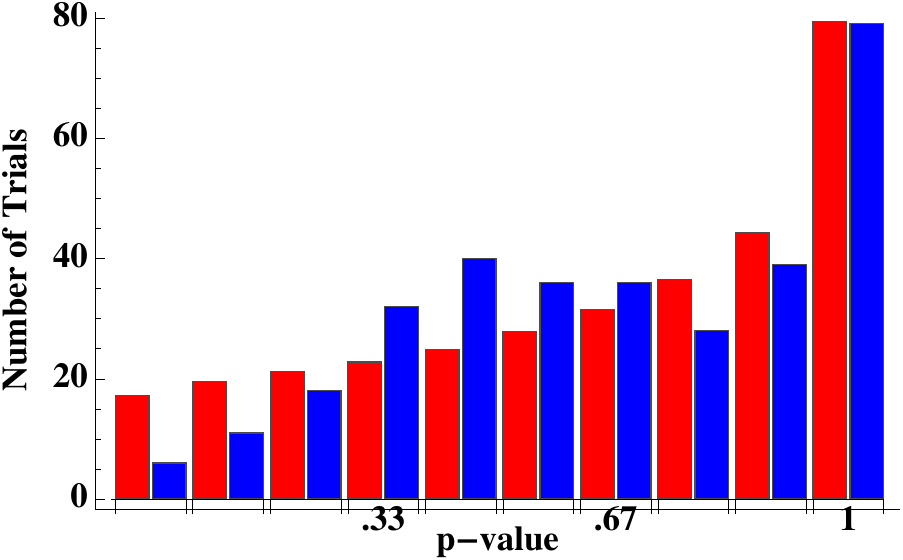}
\end{center}
\caption{Observed (blue) vs. expected (red) distribution of p-value deficits with a Gaussian error distribution for ATLAS (left) and CMS (right).}
\label{fig3}
\end{figure}

\begin{figure}[h!]
\begin{center}
\includegraphics[width=75mm]{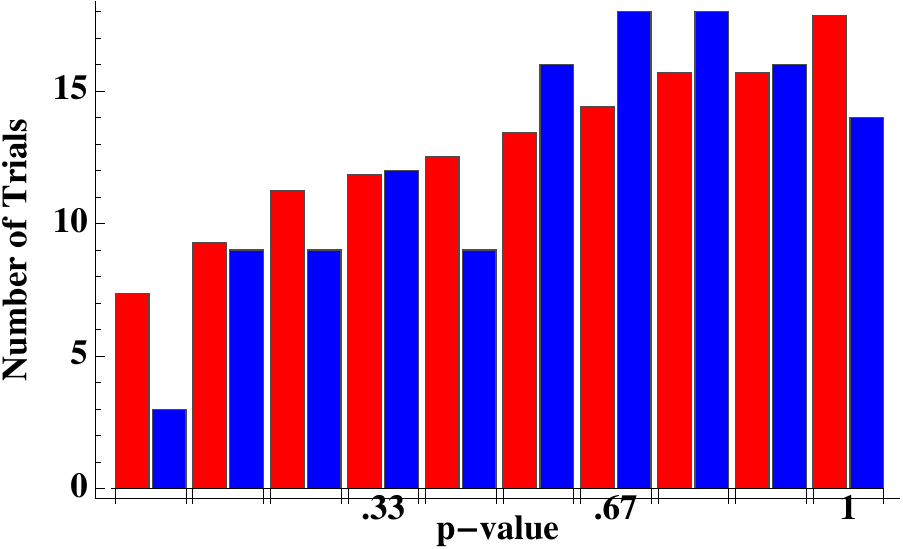}
\includegraphics[width=75mm]{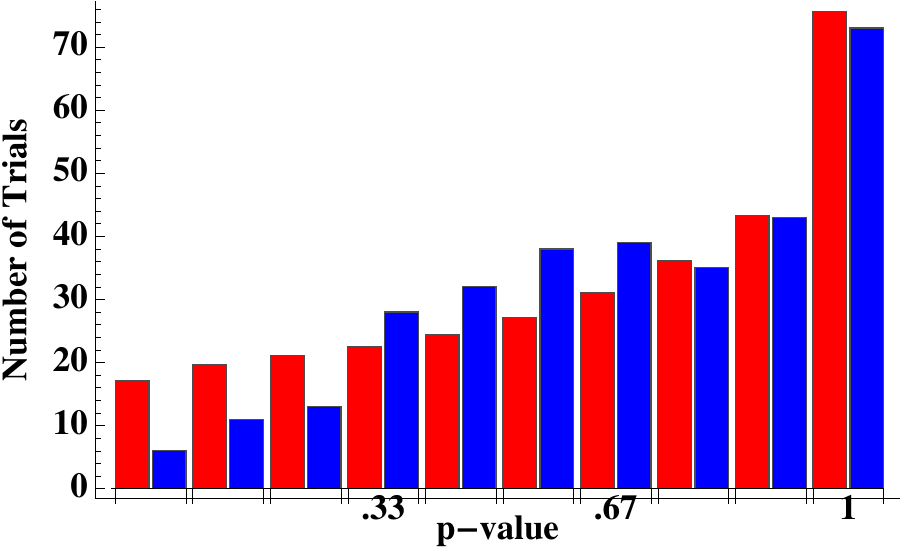}
\end{center}
\caption{Observed (blue) vs. expected (red) distribution of p-value deficits with a lognormal error distribution for ATLAS (left) and CMS (right).}
\label{fig4}
\end{figure}

The results of our analysis are shown in Figures~\ref{fig1}-\ref{fig2}.  The left-hand side of each plot represents ATLAS data, whereas the right-hand side represents CMS data.  Figures~\ref{fig1}-\ref{fig2} depict excesses above the expected signal for Gaussian and lognormal errors, respectively, whereas Figures~\ref{fig3}-\ref{fig4} depict deficits for Gaussian and lognormal errors, respectively.  Results of statistical analysis tests for the ATLAS data are shown in Tables~\ref{table1}-\ref{table2}, while CMS results are shown in Tables~\ref{table3}-\ref{table4}.

The observed distributions for the 2014 ATLAS data look very similar to those for the 2014 CMS data.  Both exhibit a lack of deficits with $p<0.1$.  Statistically, this shortage is only marginally significant for ATLAS with a p-value of $0.10$, indicating a lack of deficits at a level of $1.65\sigma$.  For CMS, on the other hand, the p-value is $0.006$, indicating a shortage of deficits at a level of $2.77\sigma$.  Furthermore, the CMS dataset displays a lack of deficits with $p<0.3$ at p-value $0.0009$, or equivalently a level of $3.32\sigma$.  The data also exhibit a statistically significant lack of p-values in the tails of the excess distribution, though this disappears for the ATLAS data once one considers deficits instead of excesses.  The greater statistical significance for the CMS distributions compared with the ATLAS distributions may be reflective of the fact that there were many more data points in the CMS dataset compared with the ATLAS dataset.

It is interesting to note that the distributions observed here are somewhat different from those observed in our analysis of the 7 TeV data~\cite{NachmanRudelius}.  That analysis also revealed a deficit of p-values in the tails of the distribution, but there were significantly fewer p-values $<0.1$, indicating a possible overestimation of the mean background as well as the uncertainty.  Here, there is actually a slight (statistically insignificant) surplus of p-value excesses $<0.1$ in the Gaussian case, but a clear lack of p-value deficits $<0.1$ in both the Gaussian and lognormal cases.  Assuming this is a result of systematic uncertainties rather than unusually large statistical fluctuations, it would indicate one of three things:
\begin{enumerate}
\item The uncertainty distributions differ significantly from Gaussian and lognormal distributions with the reported uncertainties and means.
\item The background has been underestimated as a result of biases inherent in the estimation methods.
\item The background has been underestimated as a result of new physics.
\end{enumerate}
The present analysis cannot distinguish between these three possibilities.  At the least, the differences indicate that the true uncertainty distributions are not well described by Gaussian or lognormal distributions with the reported means and uncertainties.  We therefore encourage future SUSY data searches to publish their uncertainty distributions to ensure proper interpretation of the results and more powerful analyses of the data.

\begin{table}
\begin{center}
\begin{tabular}{||c|c|c|c|c|c||} \hline
\multirow{2}{*}{Quantity} & \multirow{2}{*}{Dist. under $H_0$ ($T$)}  &\multicolumn{2}{|c|}{Test statistic ($t$)} &\multicolumn{2}{|c||}{Pr($|T|>t$)} \\\cline{3-6}
 &&Gaussian &LN&Gaussian& LN\\ \hline
Trials with $p<0.1$& N(0,1) & 0.22 & $-0.11$ &0.83& 0.91 \\ \hline
Trials with $p<0.3$&N(0,1)&0.03&0.23 &0.73& 0.82 \\ \hline
Trials with $p<0.2$ or $p>0.8$ &N(0,1)&$-2.56$& $-2.55$ & 0.01 & 0.01 \\ \hline
Expected vs. observed dist. & $\chi^2_9$& 13.65 & 17.49  & 0.14 & 0.04 \\ \hline
\end{tabular}
\end{center}
\caption{Results for statistical hypothesis tests on ATLAS excesses, under the assumptions of Gaussian and lognormal error distributions.  The distribution is significantly different from expected due to the dearth of observed p-values between $0.9$ and $1.0$.}
\label{table1}
\end{table}

\begin{table}
\begin{center}
\begin{tabular}{||c|c|c|c|c|c||} \hline
\multirow{2}{*}{Quantity} & \multirow{2}{*}{Dist. under $H_0$ ($T$)}  &\multicolumn{2}{|c|}{Test statistic ($t$)} &\multicolumn{2}{|c||}{Pr($|T|>t$)} \\\cline{3-6}
 &&Gaussian &LN&Gaussian& LN\\ \hline
Trials with $p<0.1$& N(0,1) &$ -1.66$ & $-1.65$ & 0.10 & 0.10 \\ \hline
Trials with $p<0.3$&N(0,1)&$-0.70$ & $-1.48$ & 0.49 & 0.14\\ \hline
Trials with $p<0.2$ or $p>0.8$ &N(0,1)&$-1.45$& $-1.49$ & 0.15 & 0.14\\ \hline
Expected vs. observed dist. & $\chi^2_9$& 6.23 & 6.59  &  0.72 & 0.68 \\ \hline
\end{tabular}
\end{center}
\caption{Results for statistical hypothesis tests on ATLAS deficits, under the assumptions of Gaussian and lognormal error distributions.  The lack of p-values below $0.1$ is marginally significant, while the remainder of the tests are insignificant.}
\label{table2}
\end{table}

\begin{table}
\begin{center}
\begin{tabular}{||c|c|c|c|c|c||} \hline
\multirow{2}{*}{Quantity} & \multirow{2}{*}{Dist. under $H_0$ ($T$)}  &\multicolumn{2}{|c|}{Test statistic ($t$)} &\multicolumn{2}{|c||}{Pr($|T|>t$)} \\\cline{3-6}
 &&Gaussian &LN&Gaussian& LN\\ \hline
Trials with $p<0.1$& N(0,1) & 0.13 & $-0.25$ & 0.90 & 0.80 \\ \hline
Trials with $p<0.3$&N(0,1)& $-0.56$ & 0.35 & 0.57 & 0.73 \\ \hline
Trials with $p<0.2$ or $p>0.8$ &N(0,1)& $-4.66$ & $-4.82$ & $\ll 0.001$ & $\ll 0.001$  \\ \hline
Expected vs. observed dist. & $\chi^2_9$& 28.75 & 33.33 & 0.001 &  $\ll 0.001$  \\ \hline
\end{tabular}
\end{center}
\caption{Results for statistical hypothesis tests on CMS excesses, under the assumptions of Gaussian and lognormal error distributions.}
\label{table3}
\end{table}

\begin{table}
\begin{center}
\begin{tabular}{||c|c|c|c|c|c||} \hline
\multirow{2}{*}{Quantity} & \multirow{2}{*}{Dist. under $H_0$ ($T$)}  &\multicolumn{2}{|c|}{Test statistic ($t$)} &\multicolumn{2}{|c||}{Pr($|T|>t$)} \\\cline{3-6}
 &&Gaussian &LN&Gaussian& LN\\ \hline
Trials with $p<0.1$& N(0,1) & $-2.77$ & $-2.78$ & 0.006 & 0.005 \\ \hline
Trials with $p<0.3$&N(0,1)&$-3.32$ & $-4.05$ & 0.0009 & $\ll 0.001$ \\ \hline
Trials with $p<0.2$ or $p>0.8$ &N(0,1)& $-2.81$ & $-2.55$ &  0.005 & $0.01$  \\ \hline
Expected vs. observed dist. & $\chi^2_9$& 30.09 & 24.39 &   $\ll 0.001$ & 0.003 \\ \hline
\end{tabular}
\end{center}
\caption{Results for statistical hypothesis tests on CMS deficits, under the assumptions of Gaussian and lognormal error distributions.}
\label{table4}
\end{table}

\acknowledgments

We would like to thank Lubo\v{s} Motl for his careful examination of an earlier preprint version of the analysis.  BN is supported by the NSF Graduate Research Fellowship under Grant No. DGE-4747 and also supported by the Stanford Graduate Fellowship.  TR is supported by the NSF GRF under Grant No. DGE-1144152.

\clearpage
\newpage

\bibliographystyle{JHEP-2}
\bibliography{paper}

\end{document}